\documentclass{elsart}

\usepackage{natbib,graphicx,amssymb}
\journal{New Astronomy}

\begin{document}

\begin{frontmatter}

\newcommand{\ti}{$^{44}$Ti\ }
\newcommand{\msolar}{$M_{\odot}$\,}
\newcommand{\timestento}[1]{\mbox{$\,\times\,10^{#1}$}}
\newcommand{\ttt}[1]{$\times10^{#1}$}
\newcommand{\gsim}{\mbox{%
 \raisebox{0.3ex}{$>$}\raisebox{-0.7ex}{\test*{-0.8em}$\sim$}}\/}
\newcommand{\lsim}{\mbox{\raisebox{0.3ex}{$<$}\raisebox{-0.7ex}{\hspace*{-0.8em}$\sim$}}\/}

\title{$\rm \bf ^{44}Ti$ radioactivity in young supernova remnants: Cas A and SN 1987A}

\author[1]{Yuko Motizuki \thanksref{label2}}
%\ead{motizuki@riken.jp}
and
\author[2]{Shiomi Kumagai \thanksref{label3}}
%\ead{kumagai@phys.cst.nihon-u.ac.jp}

\address[1]{RIKEN, Hirosawa 2-1, Wako, 351-0198 Japan}
\address[2]{Department of Physics, Faculty of Science and Technology,
Nihon University\\ Kanda-Surugadai 1-8, Chiyoda-ku, Tokyo 101-0062 Japan}

\thanks[label2]{E-mail: motizuki@riken.jp.
Spelling of her name (Mochizuki) has changed to Motizuki. }
\thanks[label3]{E-mail: kumagai@phys.cst.nihon-u.ac.jp}

% use the thanksref command within \title, \author or \address for footnotes:
% \title{\thanksref{label1}}
% \thanks[label1]{}
% \author{\thanksref{label2}}
% \thanks[label2]{}
% \address{\thanksref{label3}}
% \thanks[label3]{}
% including your email address
% \address{\thanksref{email}}
% \thanks[email]{E-mail: }

\begin{abstract}
We investigate radioactivity from the decay sequence of $\rm ^{44}Ti$
in young supernova remnants (SNRs), Cassiopeia~A (Cas~A) and SN~1987A.  
It is shown by a linear analysis 
that ionization of $\rm ^{44}Ti$, a pure electron capture decay isotope, 
affects the radioactivity contradistinctively in these two SNRs: 
Ionization of $\rm ^{44}Ti$ to H-like and He-like states
enhances its present radioactivity in Cas~A,
while such high-ionization decreases its radioactivity in SN~1987A.  
We briefly discuss the enhancement factor of the present radioactivity of Cas~A
considering microscopic (atomic/nuclear) physics combined with a 
hydrodynamical SNR evolution model.
For SN~1987A, we have obtained the initial $\rm ^{44}Ti$ mass of
$(0.82-2.3) \times 10^{-4} M_\odot$ from our Monte-Carlo simulations.
The resulting fluxes of $\gamma$ and hard X-rays emerged from the 
$\rm ^{44}Ti$ decay are given for the current and future experiments.
\end{abstract}


\end{frontmatter}

\section{Introduction}

The initial yield of $\rm ^{44}Ti$ that is synthesized by a single event of a 
core-collapse supernova explosion is very crucial to constrain 
dynamics of core-collapse supernova nucleosyntehsis.
This is because $\rm ^{44}Ti$ is synthesized at the vicinity of 
so-called mass cut, that divides the matter which accretes on a 
compact object and ejecta which is scattered into interstellar space.    
For this, the initial mass of $\rm ^{44}Ti$ depends sensitively on 
1) the location of the mass cut,
2) the maximum temperature and the maximum density behind the shock wave,
and 3) the internal structure 
(\mbox{\raisebox{0.3ex}{$<$}\raisebox{-0.7ex}{\hspace*{-0.8em}$\sim$}}\/
2 $\rm M_{\odot}$ from the center) of a progenitor.   

For the above reason, it is very interesting to compare theoretical
predictions of the $\rm ^{44}Ti$ yield with ``observed" values.
Since $\rm ^{44}Ti$ is radioactive, we can detect its radioactivity
and derive the initial $\rm ^{44}Ti$ mass from it; 
$\rm ^{44}Ti$ decays by electron capture to $\rm ^{44}Sc$, 
emitting 67.9 keV and 78.4 keV nuclear deexcitation lines.
Then $\rm ^{44}Sc$ decays almost exclusively by positron emission to $\rm ^{44}Ca$,
which emits 1.16 MeV deexcitation line.
The emitted positron ends up with 511 keV annihilation line. 
So far, detection of the 1.16 MeV line from Cas A 
with COMPTEL/CGRO experiment 
(e.g., Iyudin et al. 1994; Sch\"{o}nfelder et al. 2000)
and reconfirmation of this by 67.9 and 78.4 keV lines with BeppoSAX
(Vink et al. 2001) have allowed us to do such a comparison.
It is also expected to detect the $\rm ^{44}Ti$ nuclear lines 
from other young galactic SNRs and SN~1987A in LMC in near future.

The halflife of $\rm ^{44}Ti$ is a key quantity
for its radioactivity to be an important observable 
in young SNRs, and hence has been intensively studied in laboratories
after the first detection of the nuclear $\gamma$-ray flux
by Iyudin et al. (1994). 
Compilation of recent 8 experiments which were performed after
1998 (see., e.g., Hashimoto et al. 2001 and references therein;
F\"{u}l\"{o}p et al. 2000)
gives weighted mean halflife of $t_{1/2} = 60 \, \pm$ 1 yr 
(the error is 1 $\sigma$, statistical).
It is this timescale of the halflife that makes $\rm ^{44}Ti$ a useful 
diagnostic isotope.

However, a crucial point here is that $\rm ^{44}Ti$ decays only
by orbital electron capture. 
This is because the decay Q-value from the ground state of $\rm ^{44}Ti$ 
to the second excited state of $\rm ^{44}Sc$ (branching ratio of 99.3\%)
is less than twice the electron rest mass, which is at least required
for positron emission to be allowed 
by producing two 511 keV $\gamma$-photons when a positron annihilates 
with an electron (and so does that to the first excited state of 
$\rm ^{44}Sc$ for the rest of the minor fraction of the branch).
Thus we should be careful to apply the experimental halflife 
to this problem, because halflife measurements in laboratories are done 
for neutral atoms:
The electric environment for $\rm ^{44}Ti$ in a young SNR may be very much 
different from that in laboratories.  

As we shall point out later, there is a clear possibility of ionization of 
$\rm ^{44}Ti$ ongoing in SN~1987A.
Also, there are indications that $\rm ^{44}Ti$ in Cas~A is 
highly ionized; this may be confirmed directly in the future 
spectroscopic observations in X-rays. 
In this article, therefore, we are going to discuss the radioactivity of 
$\rm ^{44}Ti$ in young SNRs taking the role of ionization into consideration.
Previous studies are found in Mochizuki et al. (1999) and 
Mochizuki (2001).
In the following, a linear analysis is presented 
in section~2, and a result of the radioactivity calculated
for Cas~A is discussed in Section~3.
The current radioactivity of $\rm ^{44}Ti$ in SN~1987A is briefly
argued in Section~4.

\section{Activity Change by Ionization: A Linear Analysis}

As mentioned previously, 
the decay rate of $\rm ^{44}Ti$ depends on its electric environment.
The decay rate, $\lambda$, is proportional to the inverse of the 
halflife, 
\begin{equation}
\lambda = \frac{ln \, 2}{t_{1/2}}   .
\end{equation}
Given an ionization state, we can compute the electron-capture
rate relative to the laboratory value as precisely as we like.
However, the following approximation is good enough within the 
accuracy in question:
\begin{equation}
\lambda \approx \lambda_K +  \lambda_{LI} ,
\end{equation}
where $\lambda$ is the total decay rate, that is observed in laboratory.
The quantities $\lambda_K$ and $\lambda_{LI}$ are the partial decay rates capturing
K shell (1$\rm s_{1/2}$) and $\rm L_{I}$ shell (2$\rm s_{1/2}$) electrons, 
respectively.
The ratio of these partial decay rates for highly ionized case is given as
\begin{equation}
\frac{\lambda_{LI}}{\lambda_K} \approx \frac{1}{8}     
\end{equation}
in a nonrelativistic approximation with a point-charge field.
Adopting equations (2) and (3), we calculate the effective decay rate
for highly ionized $\rm ^{44}Ti$. 
This is shown in Table~1 relative to the laboratory value.

\begin{table}
\begin{center}
\begin{tabular}{lcccc} \hline  \hline
Ionization state & $N_e$ & $\frac{\lambda^{eff}}{\lambda}$ & $[\frac{A+\Delta A}{A}]^{\rm Cas A}$ 
& $[\frac{A+\Delta A}{A}]^{\rm 1987A}$\\  \hline  
$\rm Ti^{22+}$(fully ionized) & 0 & 0 & No activity & No activity \\
$\rm Ti^{21+}$(H-like) & 1 & 0.444 & 2.4 & 0.55 \\
$\rm Ti^{20+}$(He-like) & 2 & 0.889 & 1.3 & 0.91 \\
$\rm Ti^{19+}$(Li-like) & 3 & 0.944 & 1.2 & 0.95 \\
$\rm Ti^{28+}$(Be-like) & 4 & 1.00 & 1.0 & 1.0 \\
\hline
\end{tabular}
\caption{Change of the decay rate and the radioactivity
for highly ionized $\rm ^{44}Ti$.  
In the above,
$N_e$ is the number of bound electrons per atom, and $\lambda^{eff}$
the effective decay rate when a $\rm ^{44}Ti$ isotope has $N_e$ bound electrons.
The radioactivity including ionization effect, $(A+\Delta A)/A$, relative to that
based on the laboratory decay rate, is calculated from equation~(7)
for both Cas~A and SN~1987A.}
\end{center}
\end{table}

In Table~1, retardation of the decay is manifest in particular when
$\rm ^{44}Ti$ is in H-like and He-like ionization stages. 
Let us now consider the electron binding energies to see
if such a high-ionization is possible. 
We can simply estimate the K- and L-electron binding energies of
highly ionized atoms with
\begin{equation}
E_{e} =  \frac{(\alpha Z)^2}{2 n^2} \times 511 \, {\rm[keV]}    
\end{equation}
again under the assumption that the nucleus is a point-charge and 
the electrons are treated non-relativistically. 
In the above, $\alpha$ is the fine-structure constant, $Z$ the nuclear charge, 
$n$ the principal quantum number of an electron shell.
With equation~(4), the binding energies of K shell and $\rm L_{I}$ shell electrons
of $\rm ^{44}Ti$ for highly ionized case 
are calculated to be 6.6 keV and 1.6 keV, respectively.
It is naturally expected that 
bound electrons with the range of these binding energies can be unbound
by shock heating in SNRs seen in X-rays:
Even if the temperature of a SNR is below $E_{e}$, 
the tail of Maxwellian distribution of free electron velocity plays a role
in ionizing the elements (see Mochizuki et al. 1999).  

It should now be clear that the change of $\rm ^{44}Ti$ radioactivity 
by high-ionization in young SNRs is significant to be investigated.
The observable, radioactivity $A$, is generally expressed as
\begin{equation}
A \equiv - \frac{d N}{d t}= N_0 \, \lambda e^{- \lambda t}   .
\end{equation}
Here $N$ is the number of a radioisotope that is synthesized
in a supernova explosion, $N_0$ is its initial value, and 
$t$ is the age of a SNR.
With an observed line flux $F_{\gamma}$ the radioactivity is also 
written as 
\begin{equation}
A = \frac{4 \pi d^2 \, F_{\gamma}}{I_{\gamma} \, f_{\gamma}}  ,
\end{equation}
where $d$ is the distance to a SNR, and $I_{\gamma}$ the absolute intensity 
of the flux per decay of the parent nucleus. 
The quantity $f_{\gamma}$ is the escape fraction 
of the $\gamma$-photons, that is definitely equal to $1$ for Cas~A 
and close to $1$ for the later phase of SN~1987A.   
Combining equations (5) and (6), one can easily see that
the initial mass of $\rm ^{44}Ti$ is derived from the  
(observed) values of $F_{\gamma}$, $t_{1/2}, d$, and the age of the remnant.

Finally, a linear analysis of the radioactivity (equation~[5]) shows
\begin{equation}
\frac{\Delta A}{A} = (1 - \lambda t) \, \frac{\Delta \lambda}{\lambda} ,
\end{equation}
where $\Delta \lambda$ is the change of the decay rate
and $\Delta A$ is that of the activity.
It is worth noting that $\Delta \lambda$ is always {\em negative},
since the ionization always reduces its decay rate.
Hence the sign of $\Delta A$ is determined by that of the term 
in the parenthesis in the right side of equation~(7).
This means the following: 
If a SNR is {\em older} than the $\rm ^{44}Ti$ lifetime
(not halflife), $\sim$ 89 yrs, the activity is {\em enhanced} 
by ionization, and if {\em younger}, the activity is {\em reduced}.
Very intriguingly, we have found that the ionization phenomenon itself 
affects oppositely in Cas~A (t = 320 yrs) and SN~1987A (t = 16 yrs). 
The values of these activity changes are also shown in Table~1.  
We see in Table~1 that the radioactivity in Cas A is 
enhanced by a factor 2.4 at present and that in SN~1987A
is decreased by $\sim$45\% at present 
if all the $\rm ^{44}Ti$ atoms are in H-like ionization stage.

Actual conclusion of the effect of ionization on the radioactivity
requires the knowledge of the temperature and the density evolution of a SNR.
We are going to see that the value obtained by the linear analysis
is consistent with numerical calculations for Cas~A 
in the following session. 

\section{$\rm \bf ^{44}Ti$ Radioactivity in Cas A}

The present radioactivity is affected from the past; the point is 
whether in a SNR $\rm ^{44}Ti$ could be highly ionized and thus more
stable for a considerable period of time during the evolution.
The basic model adopted here is the same as that described in Mochizuki et al. 
(1999), in which thermal electron collisions caused by the reverse shock 
ionize $\rm ^{44}Ti$ as well as $\rm ^{56}Fe$.

We have calculated the $\rm ^{44}Ti$ radioactivity including the retardation 
of the decay as a function of time (the age of a SNR) and the position in a SNR. 
The calculations are done by solving a hydrodynamical evolution model 
(McKee and Truelove 1995) with newly introduced clumpy structure, 
combined with microscopic (nuclear/atomic) physics
where Dirac-Hartree-Slater method with finite nucleus is used to include 
electron correlations precisely.
Our model employed here is under update to include 
so-called di-electric recombination process; a free, secondary electron can be bound
to a nucleus easier when the first electron is already trapped in the orbit.
Detailed numerical calculations including this process will be presented elsewhere, 
but inclusion of this process will not change the result given below essentially.

The present result is obtained for the set of parameter values
that are consistent with recent X-ray observations of Cas~A 
(Willingale et al. 2002): 
mass of the ejecta is taken to be $2 M_\odot$ and ambient hydrogen density 
to be 15 $\rm cm^{-3}$. 
For comparison with a theoretical study of Rauscher et al. (2002), the explosion 
energy is set to be $2 \times 10^{51}$ ergs for the consistency with their 
derivation of the largest theoretical yield 
($5 \times 10^{-5} {\rm M_\odot}$) of $\rm ^{44}Ti$. 
The density enhancement factor of the clumps (Mochizuki et al. 1999) that contain 
$\rm ^{44}Ti$ is given to be $10$, which is also compatible with 
the abundance nonuniformity of the elements 
(Fe-K and Ni) reported in Willingale et al. (2001).
It has been found that relatively outer region in a SNR 
is remarkably affected by the ionization and that this region 
almost coincides with the region in Cas~A where 
Fe-K X-ray flux is observed from highly ionized Fe.
Since $\rm ^{44}Ti$ is most likely accompanied by this Fe, 
we regard the observed region of Fe-K X-rays (Willingale et al. 2001) 
as the region where $\rm ^{44}Ti$ exists in the remnant.
Averaging the radioactivity distribution in the remnant over this region, 
we obtain the averaged enhancement factor of the radioactivity
relative to that without the ionization effect 
as high as $\sim 2$ at the present age of Cas~A. 

It has been frequently argued that theoretical predictions of the initial mass 
of $\rm ^{44}Ti$ are reasonably smaller than the ``observed" initial mass
in Cas~A as inferred from the $\gamma$-line measurements on the grounds of the 
laboratory decay rate.   
Here the ``observed" initial mass is only apparent when one does not count the 
ionization effect: The real initial mass is obtained by dividing the 
``observed" mass by the averaged enhancement factor.   
Accordingly, we see that the ionization effect reduces the discrepancy between the 
``observed" value and theoretical predictions.

We found that the obtained averaged enhancement ratio of the radioactivity, 
$\sim~2$, is large enough to remove the discrepancy in    
the $\rm ^{44}Ti$ yield derived from the observed values 
and that in Rauscher et al. (2002) 
if the real flux value is located in the smaller part 
({\mbox{\raisebox{0.3ex}{$<$}\raisebox{-0.7ex}{\hspace*{-0.8em}$\sim$}}\/}
$2.5 \times 10^{-5} \, {\rm photons \, cm^{-2} \, s^{-1}}$) 
of the reported fluxes (Sch\"{o}nfelder et al. 2000 and Vink 2001)
within their uncertainties.      
However, the factor cannot compensate the discrepancy sufficiently
especially for the case that higher half
({\mbox{\raisebox{0.3ex}{$>$}\raisebox{-0.7ex}{\hspace*{-0.8em}$\sim$}}\/}
$3.5 \times 10^{-5} \, {\rm photons \, cm^{-2} \, s^{-1}}$) 
of the reported $\gamma$-ray flux (Sch\"{o}nfelder et al. 2000)
would be real.

We believe that the discrepancy is explained largely by the ionization effect.  
However, 
the remainder of the discrepancy after the subtraction of the ionization effect,
if any, may be attributed to the multi-dimensional effect of the explosion:
The production of $\rm ^{44}Ti$ in 2-D and 3-Dimensional calculations may become 
larger than that in spherical explosion models. 
The future observations of the nuclear line fluxes and the ionization 
states of $\rm ^{44}Ti$ will settle up this problem.

\section{$\rm \bf ^{44}Ti$ Radioactivity in SN 1987A}

We performed Monte-Carlo simulations of Compton degradation of the 
nuclear $\gamma$-photons emitted from the decay sequence of $\rm ^{44}Ti$ to 
explain the upper/lower bolometric luminosity observed at 3600 days after 
the explosion (Suntzeff 1997). 
Note that at the period of the observation the ionization of $\rm ^{44}Ti$
is not relevant. 
Some details of our calculation are found in Kumagai et al. (1993); 
nuclear decay parameters adopted in the present study have been updated.

We have obtained the initial $\rm ^{44}Ti$ mass of
$(0.82-2.3) \times 10^{-4} M_\odot$ 
within known accuracy of the experimental values: 
$t_{1/2} = 60 \pm 3$ yrs (3$\sigma$ deviation) and
the distance to SN~1987A, $d = 48.8 \pm 3.3$ kpc (3$\sigma$, Gould \& Uza 1998).
The expected nuclear fluxes for 6000 days after the explosion
(i.e., 2003) for both the upper 
and the lower $\rm ^{44}Ti$ masses are summarized in Table~2;
the escape fraction (equation~[6]) of the $\gamma$-ray photons, 
depending on each photon energy, 
has been taken into account here.   
Note that the derived $\rm ^{44}Ti$ mass depends on the distance 
but the expected fluxes are not.

\begin{table}
\begin{center}
\begin{tabular}{lccccc} \hline \hline
  & $\rm ^{44}Ti$ mass & 68 keV flux & 78 keV flux 
& 511 keV flux & 1.16 MeV flux \\  \hline  \hline
Upper limit & $2.3 \times 10^{-4}$ & $5.0 \times 10^{-6}$ &
$5.2 \times 10^{-6}$ & $1.0 \times 10^{-5}$ & $5.6 \times 10^{-6}$ \\ \hline 
Lower limit &$8.2 \times 10^{-5}$ & $2.5 \times 10^{-6}$ &
$2.6 \times 10^{-6}$ & $5.2 \times 10^{-6}$ & $2.8 \times 10^{-6}$ \\ \hline 
\end{tabular}
\caption{Prediction of the nuclear fluxes associated with the $\rm ^{44}Ti$ decay
in 1987A for 6000 days after the explosion (2003). 
The $\rm ^{44}Ti$ mass is given in the unit of $M_\odot$, and the fluxes 
are in photons ${\rm cm^{-2} \, s^{-1}}$.}
\end{center}
\end{table}
 
In the end, we point out that the ionization process of $\rm ^{44}Ti$ is
considered to be well underway in SN 1987A, due to
shock heating caused by the collision of the supernova blast shock with 
the dense inner ring. 
Note that H-like and He-like ionization stages of O, Ne, Mg, and 
Si have been already observed, and SN~1987A is a very rapidly evolving
remnant (see e.g., Burrows et al. 2000; Michael et al. 2002).
If $\rm ^{44}Ti$ reaches the high-ionization, the expected fluxes given
in Table~2 become smaller as discussed in the linear analysis.  
Details are found in Motizuki, Kumagai, and Nomoto (2003).

\end{document}